# Developing Artificial Mechanics Intuitions from Extremely Small Data

Jingruo Peng[1], Shuze Zhu*[1]

[1]Center for X-Mechanics, Institute of Applied Mechanics, Zhejiang University, Hangzhou 310000, China

*To whom correspondence should be addressed. E-mail: shuzezhu@zju.edu.cn

**Abstract**

Humans can possess good mechanics intuitions by learning from a few examples, which leads to the question of how to develop artificial mechanics intuitions that can be learned from small data, as we are eagerly entering the era of artificial intelligence. We propose in this Letter the sample-switchable training method, which successfully develops highly-accurate artificial mechanics intuitions that can master brachistochrone problem, catenary problem, and large nonlinear deformation problem of elastic plate by learning from no more than three samples. The model's intuitive prediction ability increases nonlinearly with respect to the number of training samples, suggesting that superb mechanics intuitions can be in-principle achieved based on a finite number of samples, reflecting how human brains form good mechanics intuitions just by learning a few cases. Our current work presents an alternative perspective for educating artificial intelligence capable of intuitively understand and predict how materials deform and move, a scenario that has been frequently seen in Science-Fiction movies.

# 1. Introduction

A well-trained structural engineer on construction sites can quickly form intuitive approximations on stress and strain state of a deforming structure, which allows quick assessment regarding the risk of structural failure. An expert of mechanics can give highly-accurate intuitive prediction on the mechanical responses of deformable structures, without performing actual calculations by writing down equations on paper or resorting to computers. These concrete examples demonstrate the advantage of mechanics intuitions in our daily lives. While intuitive predictions serve as approximations to the exact solutions with reasonable accuracy, they appear significantly faster in our minds than solving the problem on papers or in computers. Intuitive approximations could always be utilized for later refinement. Apparently, professional mechanics training is needed to possess such ability of intuitive reasoning in human brains. Entering the era of artificial intelligence, can we train artificial neural networks to possess expert mechanics intuitions?

The ultimate goal of machines is to be as intelligent as humans, as Alan Turing has argued in a seminal paper[1] published in 1950, where he raise the question "can machines think?". With the rise of computing power in recent years, there is a lot of research endeavors towards training artificial intelligence models that can match or even beat human performance using large-scale data[2–15], including large language models and generative models. Nevertheless, large-scale data model is not what we seek in this Letter. The reason is that, we prefer the way of how our human brains can form intuitions based on only a few data points[16–18]. It is well-known that human intelligence has the ability to learn quickly from limited samples and adapt to unfamiliar settings[19,20]. For instance, young children can identify new object categories given only a few examples. In the history of mechanics, Newton's Canon[21] is a famous example highlighting the power of pure thought experiment based on just a few samples. Newton imagined a cannon firing balls at different speeds on Earth. As depicted in Figure 1, sample 1 and sample 2 show that the greater the speed, the farther the ball travels, and its landing point depends on the Earth's curvature. Then, anyone who can understand the correlation between sample 1 and sample 2, would immediately acquire an intuitive generalization that with higher speed, the trajectory may match that of the Earth's surface, so that the ball will continue to fall but never reach the ground.

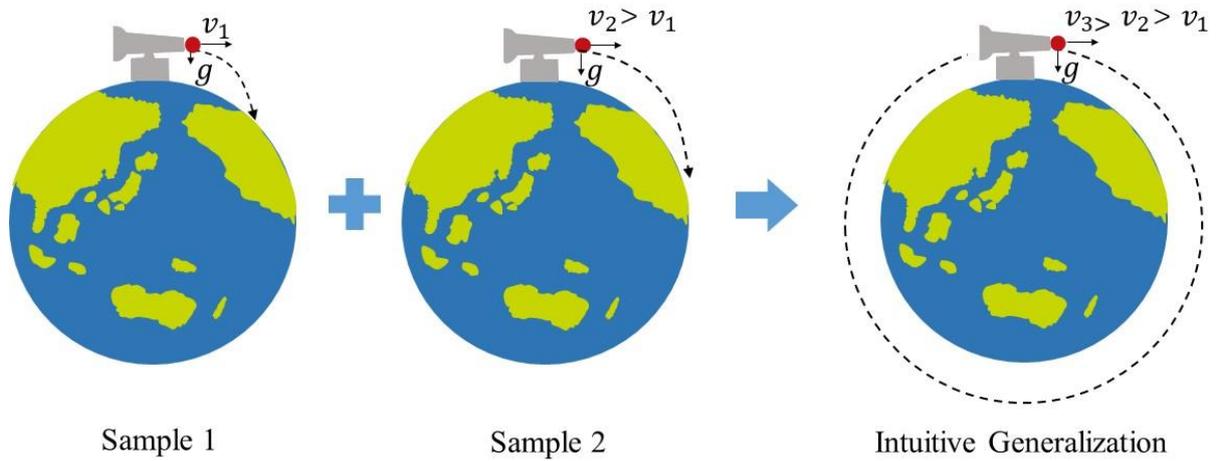

Figure 1. Newton's Cannon is a remarkable example on how intuitions can be deduced from a few samples.

The Newton's Cannon is a remarkable example on how human brains can formulate intuitions only based on a few samples. Defining mechanics intuition as the ability to guess reasonably-accurate solutions of mechanics problems of unseen samples based on a few learnt samples, the concept of training artificial mechanics intuitions may be similar with the concept of few-shot machine learning[17,22–24], although subtle differences remain. Few-shot learning problems are mainly supervised learning problem[17], as the task is to learn from a limited number of examples with supervised labeled information. Currently, few-shot learning algorithms mainly involve classification[22,25–27]. Nevertheless, labeled data, or classification tasks, are less encountered in learning to solve mechanics problems. Mechanics problem often deals with predicting field values (i.e., displacements, strains, stresses) or dynamic trajectories, which are the foundations for more complicated tasks[28–31] such as structural design[32] or topology optimization[33]. In principle, our understandings on mechanics are based on the optimization of physical functionals (e.g., the principle of minimum potential energy), and our human mechanics intuitions can be acquired by rationalizing such optimizing process qualitatively or quantitatively. Although such optimization processes rarely involve large amount of labeled data or supervised information, our human intuitions can logically correlate mechanical deformations or motion trajectories with initial conditions or boundary conditions. Can we train artificial mechanics intuitions to match or surpass human-level performance?

This Letter presents a general yet simple deep learning framework to train artificial intuition of mechanics with high-accuracy by learning from a small number of samples, simulating the experience of how humans gain mechanics intuition just from a few studying cases. The essence of current work is the method of sample-switchable training, where the loss functions are switchable among a small number of samples during training. Such treatment is analogous to how human teachers develop students' understanding on certain subject using a finite collection of exercising problems. Once the neural network is trained on a few seen samples, it is then directly applied to unseen samples for intuitive prediction (i.e., generalization to unseen samples), without further iterative process of loss function minimization. The essence of developing artificial mechanics intuitions is to optimize the accuracy when performing the above intuitive prediction over multiple unseen samples, rather than emphasizing on the prediction accuracy on a single sample as in conventional physics-guides machine learning approaches[34–38].

As demonstrations, we train artificial neural networks to gain intuitions for classical mechanics problems (the brachistochrone problem, the catenary problem) and elastic plate large-deformation problems, which are rooted deeply in the historical development of solid mechanics. We show that the trained model's intuitive prediction ability increases nonlinearly with respect to the number of samples used in training. Such a nonlinear feature suggests that excellent generalization performance can be in-principle achieved based on a finite number of samples, mimicking how human brains form good intuitions just by learning a few cases. Furthermore, our sample-switchable training method renders the neural network interpretable on the preference of loss-minimization-activity, which we called learning activity, coupled to each input features. Our current work holds potential in the future development of artificial intelligence capable of quantified intuitive prediction of how materials deform and move, a scenario that has been frequently seen in Science-Fiction movies.

## 2. Sample-Switchable Training

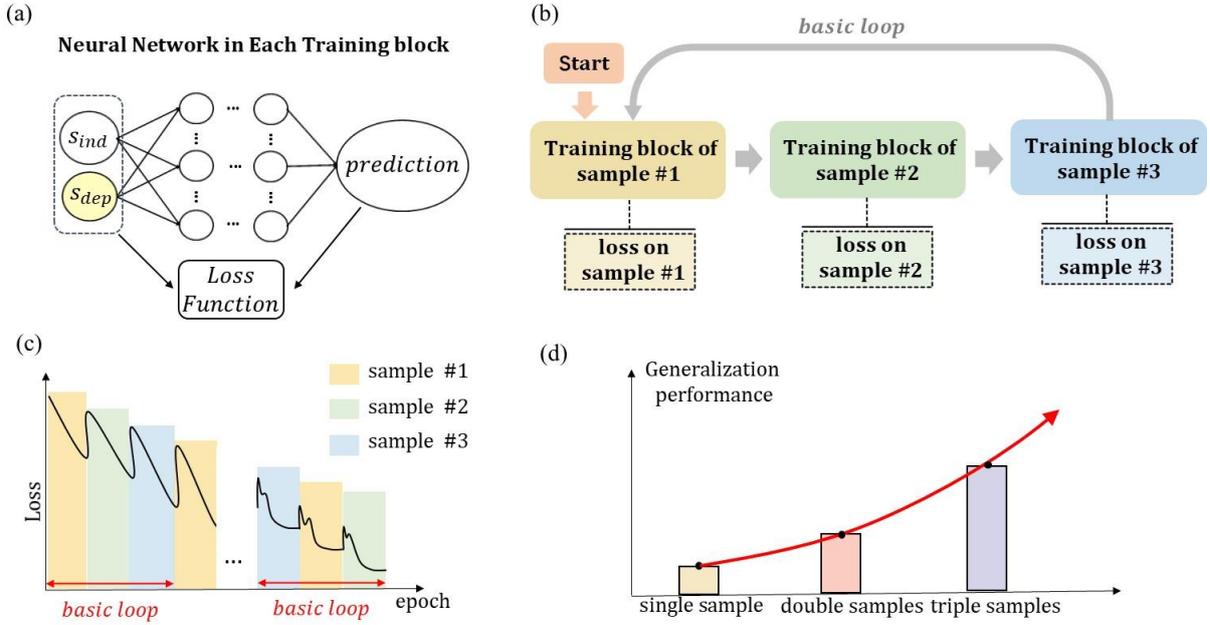

Figure 2. Illustrations on sample-switchable training. (a) The sample-dependent state feature $s_{dep}$ and the sample-independent state feature $s_{ind}$ are both explicitly encoded into the input of a Multi-Layer Perceptron (MLP), and the input of the loss function. (b) The neural network is trained against each sample cyclically (each cycle is a basic loop). At any moment the neural network only deals with the loss function for a particular sample. (c) Given the cyclically-switched training samples, evolution of training loss would exhibit sawtooth features with a decreasing overall magnitude as training continues. (d) Nonlinear enhancement in generalization performance with respect to the number of training samples.

As shown in Figure 2(a), a particular feature of our neural network architecture is that the sample-dependent state feature (e.g., the variable $s_{dep}$) along with the sample-independent state feature (e.g., the variable $s_{ind}$) are both explicitly encoded into the input of a Multi-Layer Perceptron (MLP), and the input of the loss function. The loss functions to be minimized encode the physical functionals to be optimized given the physical input features.

Given the separation of sample-dependent state features with the sample-independent state features, when multiple samples are sequentially and cyclically used in training, the neural network will prioritize on learning the hidden significance of sample-dependent feature to minimize the sample-dependent loss function, therefore gaining intuitions to produce results that probably minimize the loss functions that are coupled with unseen sample.

In current work, the input of the neural network also serves as the only input of the loss function for each sample, and at any moment the neural network only deals with the loss function for a particular sample. During sample-switchable training, the number of epochs for training a single sample is fixed. If such number is reached, both the input and the loss function of the neural network are switched to next sample, while current set of neural network parameters is inherited (Figure 2b). In other words, our neural network is trained against each sample cyclically while taking advantage of the understanding on previous training samples. We call each cycle as a basic loop (Figure 2b). Upon switching to the next sample, the neural network must learn new dependencies on the sample-dependent state so that the loss suddenly increases. The idea is that, we deliberately enforce the artificial neural network to focus on learning the influence of sample-dependent features by repeating the basic loops. Given the cyclically-switched loss function, it is natural to expect that the evolution of training loss would exhibit sawtooth features (Figure 2c) with a decreasing overall magnitude as training continues.

Once the overall magnitude of loss function reaches a stable phase, its generalization performance (intuitive prediction ability) is tested, which is measured on the closeness to the exact solution (i.e., the ground truth). During such test, the learned neural network is directly applied to unseen samples, without further iterative process of loss minimization. Because the sample-dependent feature is an explicit input to the MLP, one can easily utilize the trained MLP to generate (or guess) predictions on a wide range of unseen samples. Our results show that there is a nonlinearly enhanced generalization performance (Figure 2d) with respect to the number of samples used in training. In current study, the maximum number of samples used for training is 3.

Our sample-switchable training shares some similarities with stochastic gradient descent supervised learning of labeled samples in the sense that a small subset of samples is randomly selected for training. However, stochastic gradient descent supervised training is a purely data-driven approach that does not emphasize on physical correlation from sample-dependent input to output. In contrast, our network is forced to learn to optimize physical functional values (not supervised learning task) by minimizing the perturbative errors during constantly-switched sample-dependent inputs, which is crucial for obtaining causal intuitions. Therefore, for machine learning tasks that involves optimizing physical functional values, if the input can be partitioned into sample-dependent and sample-independent components, our proposed sample-switchable training method can be applied. In this sense, our method is compatible with a wide range of artificial neural network architecture and high-dimensional problems.

Physics-informed neural network (PINN)[35,37] is typically used to solve physical problems on a physical domain of interests. An important property of PINN is that a small amount of supervised training data (allocated points of known boundary conditions) plus unsupervised allocation points (to-be-solved allocated points whose field values must satisfy governing equation) is sufficient to predict the full-field solution. However, the typical input features of PINN only contain sample-independent input features in our definition (i.e., the coordinates of allocated points[35,37]) so that conventional PINN cannot treat another unseen sample without re-training the neural network parameters if the boundary conditions are changed. In contrast, our work emphasizes more on the small number of samples (e.g., each sample is an individual problem that can be solved by conventional PINN) and the generalization performance for unseen samples without re-training the neural network parameters (e.g., there are sample-dependent input features). We aim to train a network that can generalize well over a wider range of unseen samples, rather than the accuracy of predicted field values over a single sample. In order to enable the cross-sample generalization, the sample-dependent feature (e.g., the height different in brachistochrone problem) is explicitly encoded in the neural network input, which is a key feature of sample-switchable training method.

## 3. Intuitional learning on trajectories or shapes in two dimensions

We begin with the demonstrations in learning intuitions over trajectories or shapes in two dimensions. We focus on brachistochrone problem and catenary problem, which represent classic mechanics problems without and with constraints.

### 3.1 The brachistochrone problem

The brachistochrone problem (Figure 3(a)) may be arguably the most famous problem in the history of mechanics. Consider a trajectory that connects two points A and B on a vertical plane, which are not on the same vertical line. The objective is to find the trajectory that corresponds to the shortest time with which a particle released at A moves along the trajectory towards B only by the effect of gravity. See Section 1.1 in Supporting Information (SI) for detailed description on the mathematical models.

The input of the network includes the x-coordinate, which is sample-independent, and the height difference $h$ between point A and point B, which is sample-dependent. For a given sample, the final output is the y-coordinate at the input x-coordinate. Specifically, we construct a MLP consisting of two hidden layers, each with five neurons (Figure 3(b)), so that the output of the MLP is denoted as $y(x, h) = MLP(x, h)$, which corresponds to a trajectory with a particular travel time. The loss function of the neural network is the above trajectory-coupled travel time plus boundary penalized terms (i.e., in order to impose constraint that the trajectory passes point A and point B), so that the minimization of loss function during training generally produces the fastest trajectory. Symbolically, the loss function is denoted as $L(x_1, x_2, \ldots, x_{Q_P}, MLP(x_1, h), MLP(x_2, h), \ldots, MLP(x_{Q_P}, h), h)$, where $Q_P$ is the number of interpolating points.

For demonstration, we consider a problem where the horizontal distance between two ends point is $30\pi$. Three different samples with $h =$ 30, 50 and 70 are considered. Next, the trajectory is discretized by 11 interpolating points (including point A and point B). Three single-sample networks (labeled as $h$30, $h$50, $h$70) are trained only using single sample. Two double-sample networks (labeled as $h$(30-50), $h$(50-70)) are trained using double samples. Another triple-sample network (labeled as $h$(30-50-70)) is trained using all three samples. See Section 1.2 in SI for detailed description on the implementation of neural network.

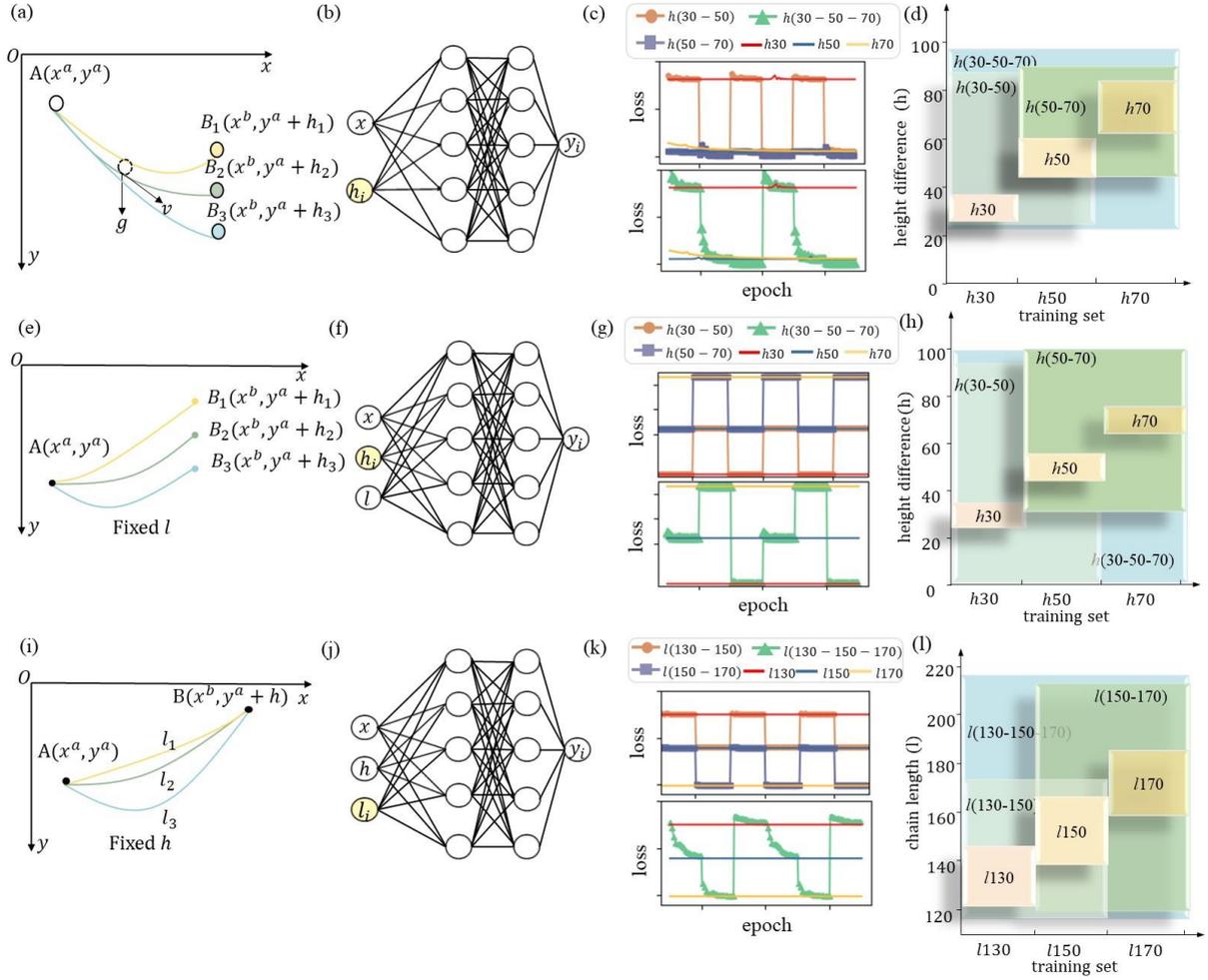

Figure 3. The schematics (a, e, i), neural network configurations (b, f, j), evolution of training loss at stable phase (c, g, k), and the generalization performance (d, h, l) on the brachistochrone problem, the catenary problem with the same height difference and the same chain length. In (a), the $i$th sample has height difference $h_i$, which is the sample-dependent input in (b) (colored in yellow) during sample-switchable training. (c) As the training loss reaches stable phase, sample-dependent training loss value appears during sample-switching. The vertical spans of the colorful blocks in (d) describe the generalization ranges where the $R^2$ to the exact solution exceed 0.9. The horizontal spans of the colorful blocks in (d) represent the samples used in training, whose corresponding model names are marked on the blocks. The bottom two rows show the corresponding results for the catenary problem trained on samples of the same chain length $l$ but distinct height difference $h_i$ (e-h), and on samples of the same height difference $h$ but distinct chain length $l_i$ (i-l).

During the sample-switchable training, we monitor the evolution of loss of all six models, which gradually decreases with increasing epochs, and eventually the overall magnitude of loss functions is minimized to a stable phase (see Section 1.3 in SI). Upon sample switching during the basic loop, the sudden loss jump appears as expected, which also continues into the stable phase (Figure 3(c)). It is important to realize that the physical meaning of the loss value is related to the travel time, so that as the training converges to stable phase, we can see that the loss of the model $h(30\text{-}50\text{-}70)$ is constantly switched among the loss associated with single sample (note that the stable loss values for $h50$ and $h70$ appear to have similar magnitude).

To test the generalization performance of each trained model, height difference $h$ is fed into $y(x, h) = MLP(x, h)$, which immediately renders guessed optimal trajectory coupled to each $h$. For a given $h$, if the correlation coefficient $R^2$ between the guessed trajectory and the exact solution exceeds 90%, then the current $h$ is within the generalization range of current model. In other words, the model has gained rather accurate intuitions over the solutions coupled to current $h$. Figure 3(d) summarizes the generalization ranges for all models, whose names are marked in corresponding blocks. The vertical spans of blocks describe the generalization range, and the horizontal spans of blocks describes the group of samples used in training.

The generalization ranges from the triple-sample model and double-sample model are generally much larger than the simple superposition of those from single-sample model. For example, the $h$30 model can generalize for the height difference between 26 and 37. The $h$50 model can generalize for the height difference between 44 and 61. However, the upper and lower limit of generalization range of $h$(30-50) model cover the simple union of the individual generalization ranges from $h$30 model and $h$50 model. The generalization range of the $h$(30-50-70) model can cover the height difference reaching 100. Section 1.4 in SI illustrates the generalization performance by visually showing the predicted trajectories. Section 1.5 in SI illustrates distribution of the $R^2$ scores of all models when the generalization performance is tested on a wide range of sample-dependent input values.

Our sampling switching method with sample-dependent input features can even learn good intuitions from highly-similar samples. For example, consider a brachistochrone problem model $h$(29-30-31). To convincingly reveal the evolving quality of generalization, we compare the models' generalization performance ($R^2 >= 0.90$) at a wide range of training epochs when the training is considered converged. Particularly, as shown in Section 1.6 in SI, when both models are converged after iterations, the sample-switchable model of $h$(29-30-31) has a much broader continuous generalization range (e.g., h range [17, 76]) than the union set of generalization ranges from the single-sample models (e.g., h range [26, 40]). This clear contrast of generalization performance supports our claim that sample-switchable training with sample-dependent input features is advantageous for learning mechanics intuitions.

## 3.2 The catenary problem

Next, we discuss the catenary problem (Figure 3(e), 3(i)), where a perfectly flexible and uniform chain of length $l$, with constant mass per unit length, is suspended between two fixed ending points A and B with height difference $h$. The objective is to determine the chain shape under gravity. We analyze the problem in two categories. In the first category, $l$ is held constant and we train mechanics intuitions over varying $h$. In the second category, $h$ is fixed and we train mechanics intuitions over varying $l$.

The basic machine learning settings are similar with those from the previous brachistochrone problem, except that the catenary problem necessitates the consideration of constraint on the chain length, which is addressed by implementing a penalty function (see Section 2.1 in SI for detailed description on the mathematical models). Specifically, we construct a MLP consisting of two hidden layers (Figure 3(f), 3(j)), each with five neurons. The output of the MLP for catenary problem is denoted as $y(x, h, l) = MLP(x, h, l)$, where the chain length $l$ and the height difference $h$ can be sample-dependent input features. The loss function includes the gravity potential energy, the boundary penalty, plus the chain length penalty. Symbolically, the loss function is denoted as $L(x_1, x_2, ..., x_{Q_P}, MLP(x_1, h, l), MLP(x_2, h, l), ..., MLP(x_{Q_P}, h, l), h, l)$, where $Q_P$ is the number of interpolating points. For demonstration, we consider the case where the horizontal distance between two ending points is $30\pi$. See Section 2.2 in SI for detailed description on implementation of neural network.

In the first category (Figure 3(e)), all training samples have the same chain length $l$=140, yet they differ in ending point height difference of $h$=30, 50, 70. Then networks, labeled as $h$30, $h$50, $h$70, are trained using single samples. Networks, labeled as $h$(30-50), $h$(50-70), $h$(30-50-70), are trained using hybrid samples. In the second category (Figure 3(i)), all training samples have the same height difference $h$=50, yet they differ in chain length of $l$=130, 150, and 170. Then networks, labeled as $l$130, $l$150, $l$170, are trained using single samples. Networks, labeled as $l$(130-150), $l$(150-170), $l$(130-150-170), are trained using hybrid samples. The actual number of training samples is evident in the model labels.

It is important to realize that the physical meaning of the loss value is ultimately related to gravity potential, so that as the training converges to stable phase, the loss of the model trained with hybrid samples is constantly switched among the loss associated with single sample (e.g., model $h$(30-50-70) in Figure 3(g), model $l$(130-150-170) in Figure 3(k)). After the loss function values in all trained networks reach steady state (see Section 2.3 in SI), we then test the generalization performance.

To test the generalization performance over height difference, $h$ is fed into the trained $y(x, h, l) = MLP(x, h, l)$ of the first category. To test the generalization performance over chain length, $l$ is fed into the trained $y(x, h, l) = MLP(x, h, l)$ of the second category. The interval within which $R^2$ to exact solution exceeds 90% is defined as the generalization range that intuitive prediction is acceptable. As generally shown in Figure 3(h) and Figure 3(l), we similarly see that the predictive ranges from the triple-sample model and double-sample model are much larger than the simple superposition of those from single-sample model. Section 2.4 in SI visualizes the generalization performance using predicted trajectories.

## 4. Prioritized learning on the hidden significance of sample-dependent features

The advantage of our proposed sample-switchable training is the clear interpretability on the logics on how the "intuitions" are formed in neural network training. In this section, using brachistochrone problem and catenary problem as examples, we show how the sample-switchable training can be rationalized and interpreted. By analyzing the learning activity along all the neuron paths connecting input features to output values, we can identify which input features are considered the most critical by the model, because the model spends the largest amount of effort trying to minimize the loss due to varying input features.

A learning path is defined as a sequence of neuron connection segments that bridges the input neuron to the output neuron. We define a learning activity coupled to each neuron connection segment, which can be calculated using gradient of the loss function with respect to the input features along learning path based on the chain rule as follows. Assuming that during training, a total of $n$ interpolating points (e.g., see the definition in previous Section 3.1) are used to calculate the loss. Let's consider the learning path connecting the $i_0^{th}$ neuron in the $Z_0$ layer (the input layer), the $i_1^{th}$ neuron in the $Z_1$ layer (the first hidden layer), the $i_2^{th}$ neuron in the $Z_2$ layer (the second hidden layer), and the $i_3^{th}$ neuron in the $Z_3$ layer (the output layer) at epoch $t$. For interpolating point $j$, such learning path is labeled as $(i_0 i_1 i_2 i_3)^j$, and neurons along the path are sequentially labeled as $z_0^{i_0 j}$, $z_1^{i_1 j}$, $z_2^{i_2 j}$, $z_3^{i_3 j}$. The gradient of the total loss with respect to the input neuron $z_0^{i_0 j}$ can be expressed as $\frac{\partial loss}{\partial z_0^{i_0 j}} = \sum_{i_1, i_2, i_3} \frac{\partial loss}{\partial z_3^{i_3 j}} \frac{\partial z_3^{i_3 j}}{\partial z_2^{i_2 j}} \frac{\partial z_2^{i_2 j}}{\partial z_1^{i_1 j}} \frac{\partial z_1^{i_1 j}}{\partial z_0^{i_0 j}}$. Then the measure of learning activity of path $(i_0 i_1 i_2 i_3)^{total}$ considering all interpolating points at epoch $t$ is calculated as $\sum_{j=1}^{n} \left| \frac{\partial loss}{\partial z_3^{i_3 j}} \frac{\partial z_3^{i_3 j}}{\partial z_2^{i_2 j}} \frac{\partial z_2^{i_2 j}}{\partial z_1^{i_1 j}} \frac{\partial z_1^{i_1 j}}{\partial z_0^{i_0 j}} \right|$. The average learning activity $LA(i_0 i_1 i_2 i_3)$ along the path $(i_0 i_1 i_2 i_3)^{total}$ from the start of training to epoch $t$ (at which the model converges) can be expressed as $LA(i_0 i_1 i_2 i_3) = \frac{1}{t} \sum_t \left( \sum_{j=1}^{n} \left| \frac{\partial loss}{\partial z_3^{i_3 j}} \frac{\partial z_3^{i_3 j}}{\partial z_2^{i_2 j}} \frac{\partial z_2^{i_2 j}}{\partial z_1^{i_1 j}} \frac{\partial z_1^{i_1 j}}{\partial z_0^{i_0 j}} \right| \right)$. See Section 3 in SI for visual illustrations.

The $LA(i_0 i_1 i_2 i_3)$ are colored according to their magnitude to reveal the most learning-active neuron connection pathways between the input layer and the output layer. The learning paths corresponding to top values are visualized with bold lines. In the brachistochrone problem (Figure 4(a-c)), the sample-dependent feature $h$ is found to possess most learning-activate neuron connections. Similarly, in the catenary problem with varying height differences (Figure 4(d-f)), we find that most of the largest $LA(i_0 i_1 i_2 i_3)$ belong to the neuron connections to the sample-dependent feature $h$. In the catenary problem with varying rope lengths (Figure 4(g-i)), we find that most of the largest $LA(i_0 i_1 i_2 i_3)$ belong to the neuron connections to the sample-dependent feature $l$. In contrast, in single-sample training for both two problems, where the input features are fixed, the sample-independent features are found to possess the most learning-activate neuron connections during the training process (see Section 3 in SI).

The above analysis shows that, in sample-switchable training, the neural network strives to learn the impact of varying sample-dependent input features from a small set of samples, which is similar with how our human brains understand key influencing factors to solve problems. This mechanism significantly improves the generalization performance of neural networks, particularly on small-sample datasets.

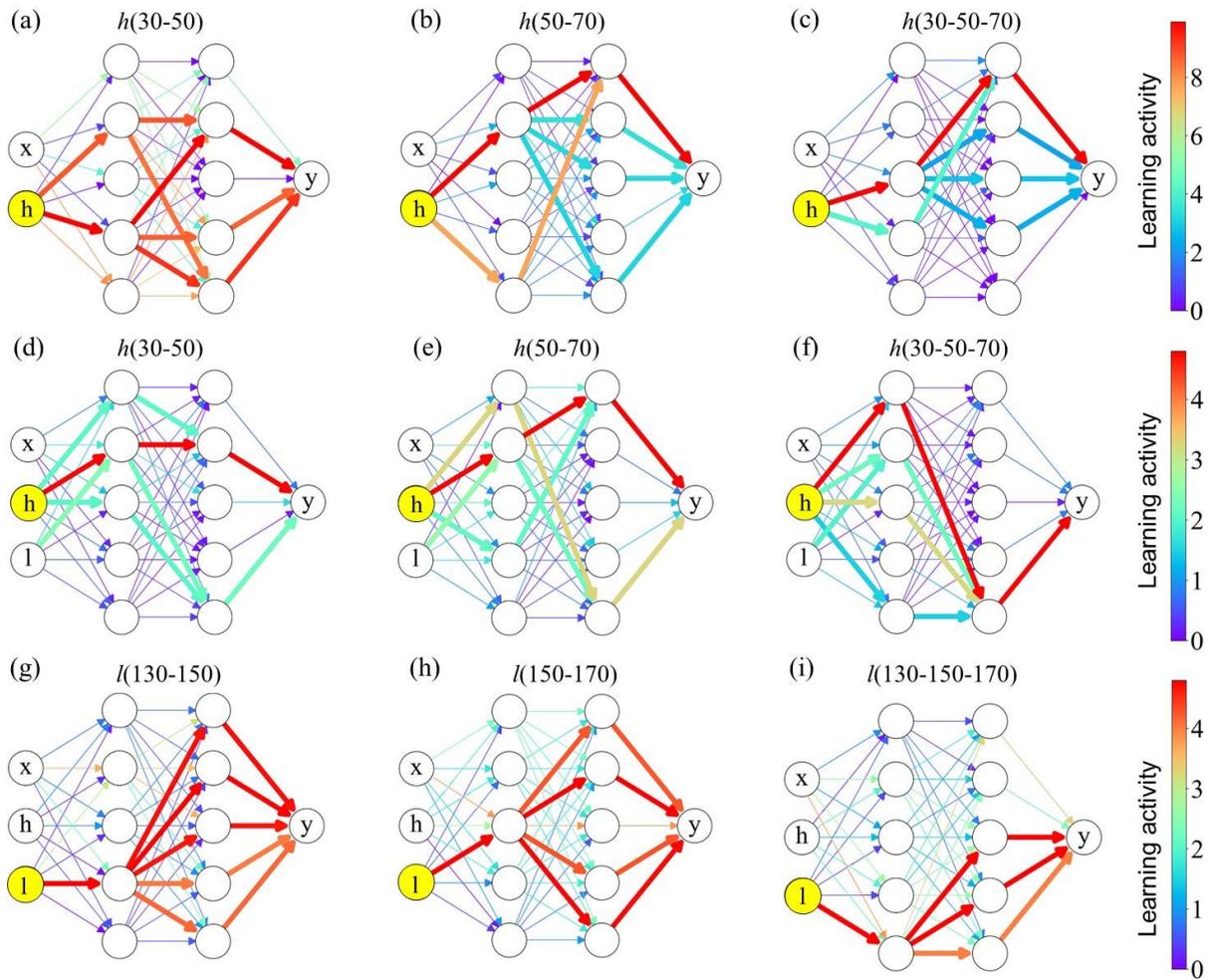

Figure 4. Demonstration on the interpretability of sample-switchable training method. The neural pathways originating from the sample-dependent input features (circled in light yellow) exhibit the highest learning activity. Each panel is labeled with the model's name. (a-c) are for brachistochrone problem, and (d-i) are for catenary problem.

## 5. Intuitional learning on displacement fields in three dimensions

In this section, we demonstrate the application of our method to the large deformation problem in elastic plate, with the goal to guess displacement fields with acceptable accuracy. The elastic plate has a square shape, with clamped boundary condition. There is a uniformly distributed pressure applied perpendicular to the plate's surface. The aim is to guess the in-plane and out-of-plane displacement fields. Geometric nonlinear strain is considered (see Section 4.1 in SI for details). We plan to train intuitions for varying pressure (Figure 5(a-d)), varying plate size (Figure 5(e-h)), and varying flake orientation to test if rotational transformation is understood (Figure 5(i-l)).

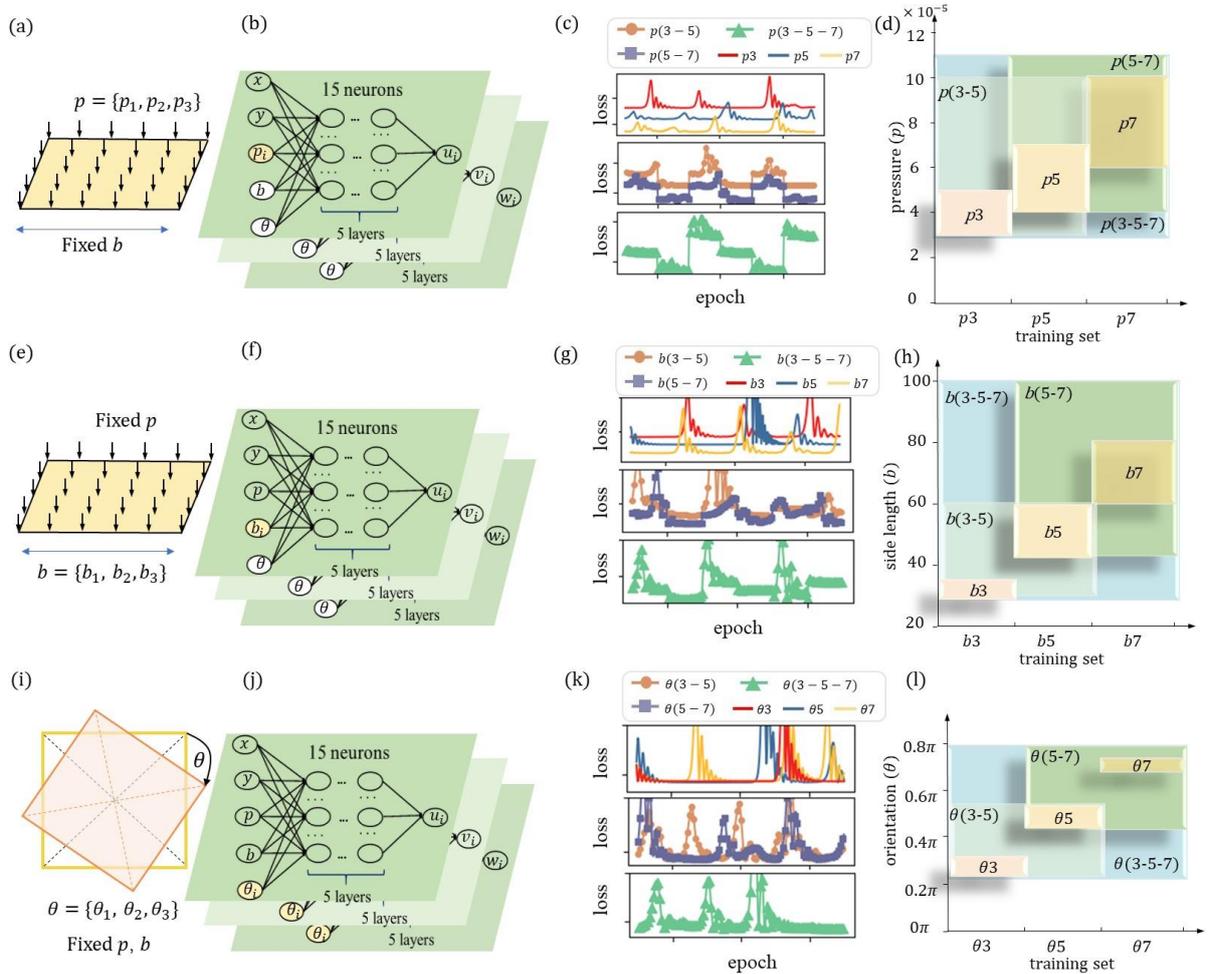

Figure 5. The schematics (a, e, i), neural network configurations (b, f, j, showing for each displacement there is an independent MLP of 5 hidden layers, each with 15 neurons), evolution of training loss at stable phase (c, g, k), and the generalization performance (d, h, l) on the elastic plate large deformation problem. In (a), the $i$th sample is subjected to external pressure $p_i$, which is the sample-dependent input (colored in yellow) in (b) during sample-switchable training. As the training becomes stable, sample-dependent training loss value appears in (c). The vertical spans of the colorful blocks in (d) describe the generalization ranges where the $R^2$ to the exact solutions for all displacement fields exceed 0.9. The horizontal spans of the colorful blocks in (d) describe the samples used in training, whose corresponding model names are marked on the blocks. The bottom two rows show the corresponding results for the elastic plate problem trained on samples of the same orientation but distinct side lengths $b_i$ (e-h), and on samples of the same side length $b$ but distinct orientation $\theta_i$ (i-l).

Mathematically, the strain energy of the elastic plate (see Section 4.1 in SI for details) is considered as $U = \frac{Eh}{2(1-\mu^2)} \iint \left\{ \left(\frac{\partial u}{\partial x}\right)^2 + \frac{\partial u}{\partial x}\left(\frac{\partial w}{\partial x}\right)^2 + \left(\frac{\partial v}{\partial y}\right)^2 + \frac{\partial v}{\partial y}\left(\frac{\partial w}{\partial y}\right)^2 + \frac{1}{4}\left[\left(\frac{\partial w}{\partial x}\right)^2 + \left(\frac{\partial w}{\partial y}\right)^2\right]^2 + 2\mu\left[\frac{\partial u}{\partial x}\frac{\partial v}{\partial y} + \frac{1}{2}\frac{\partial v}{\partial y}\left(\frac{\partial w}{\partial x}\right)^2 + \frac{1}{2}\frac{\partial u}{\partial x}\left(\frac{\partial w}{\partial y}\right)^2\right] + \frac{1-\mu}{2}\left[\left(\frac{\partial u}{\partial y}\right)^2 + 2\frac{\partial u}{\partial y}\frac{\partial v}{\partial x} + \left(\frac{\partial v}{\partial x}\right)^2 + 2\frac{\partial u}{\partial y}\frac{\partial w}{\partial x}\frac{\partial w}{\partial y} + 2\frac{\partial v}{\partial x}\frac{\partial w}{\partial x}\frac{\partial w}{\partial y}\right]\right\}dxdy + \frac{Eh^3}{24(1-\mu^2)} \iint \left\{ \left(\frac{\partial^2 w}{\partial x^2} + \frac{\partial^2 w}{\partial y^2}\right)^2 + 2(1-\mu)\left[\left(\frac{\partial^2 w}{\partial x \partial y}\right)^2 - \frac{\partial^2 w}{\partial x^2}\frac{\partial^2 w}{\partial y^2}\right]\right\}dxdy$, where $u, v$ are two in-plane displacements, and $w$ is the out-of-plane displacements. The boundary conditions for the elastic plate are $u = 0, v = 0, w = 0$. There is external work ($p$ denotes external distributed load) $W = \iint (w * p) dxdy$, then the total potential energy to be minimized is $E = U - W$.

The MLP for each displacement field comprises five hidden layers, each with 15 neurons (Figure 5(b), 5(f), 5(j)). See Section 4.2 in SI for detailed description on implementation of neural network. The clamped boundary conditions can be handled by directly applying a clamp operation to the network's output. To calculate the total potential energy, the elastic plate is randomly discretized into $Q_p$ points for given orientation. The inputs to the neural network include the x-y-coordinates of the above discretized points, the magnitude of the uniformly distributed load $p$, the length of the elastic plate's sides $b$, and the orientation angle $\theta$. Symbolically, the outputs of the MLP for elastic plate problem can be denoted as $u(x, y, p, b, \theta) = MLP_u(x, y, p, b, \theta)$, $v(x, y, p, b, \theta) = MLP_v(x, y, p, b, \theta)$, and $w(x, y, p, b, \theta) = MLP_w(x, y, p, b, \theta)$. Note there each displacement field has independent MLP parameters from the same MLP network structure (Figure 5b, 5f, 5j). The loss function includes the potential energy and boundary penalty, which can be symbolically denoted as

$L(x_1, x_2, \ldots, x_{Q_p}, y_1, y_2, \ldots, y_{Q_p}, u(x_1, y_1, p, b, \theta), \ldots, u(x_{Q_p}, y_{Q_p}, p, b, \theta), v(x_1, y_1, p, b, \theta), \ldots, v(x_{Q_p}, y_{Q_p}, p, b, \theta), w(x_1, y_1, p, b, \theta), \ldots, w(x_{Q_p}, y_{Q_p}, p, b, \theta), p, b, \theta)$. Again, we emphasize that the input of loss function completely comes from the input of the neural network.

We consider formulating training samples with a range of side lengths (units: mm), orientations (units: radian), and pressures (units: MPa). See Section 4.1 in SI for details of material parameters. To develop the intuition over varying pressure $p$, we prepare samples with fixed side length ($b = 30$) and fixed orientation ($\theta = 0$). Networks, labeled as $p3, p5, p7$, are each trained using single pressure (i.e., model $p3$ stands for pressure $p = 3 \times 10^{-5}$). Networks, labeled as $p(3-5), p(5-7), p(3-5-7)$, are each trained using multiple pressures (i.e., model $p(5-7)$ uses pressure $p = 5 \times 10^{-5}$ and $p = 7 \times 10^{-5}$ for switchable training). To develop the intuition over varying side length $b$, we prepare samples with fixed pressure ($p = 1 \times 10^{-5}$) and fixed orientation ($\theta = 0$). Networks, labeled as $b3$, $b5$, $b7$, are each trained using single side length (i.e., $b3$ stands for side length $b = 30$). Networks, labeled as $b(3\text{-}5)$, $b(5\text{-}7)$, $b(3\text{-}5\text{-}7)$, are each trained using multiple side lengths (i.e., $b(5\text{-}7)$ uses side length $b = 50$ and $b = 70$ for switchable training). To develop intuitions over varying orientation $\theta$, we prepare samples with fixed pressure ($p = 1 \times 10^{-5}$) and fixed side length ($b = 30$). Then networks, labeled as $\theta3$, $\theta5$, $\theta7$, are each trained using single orientation (i.e., $\theta3$ stands for $\theta = 0.3\pi$). Networks, labeled as $\theta(3\text{-}5)$, $\theta(5\text{-}7)$, $\theta(3\text{-}5\text{-}7)$ are each trained using multiple pressure (i.e., $\theta(5\text{-}7)$ uses $\theta = 0.5\pi$ and $\theta = 0.7\pi$ for switchable training).

Again, the generalization performance is assessed after the training loss is considered to have converged to the stable phase. Since the physical meaning of the loss value is majorly related to potential energy of each sample, the stable phase ensures that the potential energies for all samples are considered close to minima where the solutions are found. Therefore (see Section 4.3 in SI), within the stable phase, the loss of the model trained with hybrid samples is constantly switched among the loss associated with single sample (e.g., model $p(3\text{-}5\text{-}7)$ in Figure 5(c), model $b(3\text{-}5\text{-}7)$ in Figure 5(g), model $\theta(3\text{-}5\text{-}7)$ in Figure 5(k)), which could be understood in that the potential energies are sample-dependent. Considering all cases (Figure 5(d), 5(h), 5(l)), generalization ranges again demonstrate a steady enhancement on intuitive prediction range with respect to the number of samples used for training. Section 4.4 in SI illustrates the generalization performance by visually showing the predicted displacement fields for all three categories.

## 6. Conclusions

In this work, we develop artificial yet excellent mechanics intuitions from extremely-small data, using brachistochrone problem, the catenary problem, and large deformation problem of elastic plate as demonstrations. The proposed sample-switchable training method that has advantages of good interpretability, and great extensibility over the number of samples. We show that the trained network from sample-switchable method can effectively learn the correlation between the predicted quantities and the sample-dependent features (e.g., boundary conditions), leading to much enhanced generalization performance. Our trained neural network appears to be able to "think" like a human, quickly identifying the mechanisms of sample-dependent features, and can give intuitive predictions whose accuracy is at least comparable to human comprehension.

Our current work may inspire the development of few-shot learning protocols on educating artificial intelligence for mechanics.